\documentclass[balance,upint,subscriptcorrection,varvw,mathalfa=cal=boondoxo,spanish,french,vietnamese,russian,greek,pdf-a,colorlinks]{asmeconf}

\usepackage{array}
\usepackage{booktabs}
\usepackage{colortbl}
\usepackage{makecell}
\usepackage{tabularx}
\usepackage{multirow}
\usepackage{microtype}
\usepackage{soul}
\usepackage{xcolor} 
\definecolor{lightyellow}{rgb}{1.0, 1.0, 0.88} 
\sethlcolor{lightyellow} 

\hypersetup{%
	pdfauthor={John H. Lienhard},									  
	pdftitle={ASME Conference Paper LaTeX Template},                  
	pdfkeywords={ASME conference paper, LaTeX template, BibTeX style},
	pdfsubject = {Describes the asmeconf LaTeX template},			  
	pdflicenseurl={https://ctan.org/pkg/asmeconf},
}

\begin{document}
\begin{sloppypar}

\ConfName{Proceedings of the ASME 2024\linebreak International Design Engineering Technical Conference and\linebreak Computers and Information in Engineering Conference}
\ConfAcronym{IDETC-CIE2024}
\ConfDate{August 25-28, 2024} 
\ConfCity{Washington, DC, USA} 
\PaperNo{DETC2024-143163}


\title{TRIZ-GPT: An LLM-augmented method for problem-solving} 
 
%
%
%

\SetAuthors{%
	Liuqing Chen\affil{1}, 
	Yaxuan Song\affil{1}, 
	Shixian Ding\affil{1}, 
        Lingyun Sun\affil{1},  
	Peter Childs\affil{2},  
	Haoyu Zuo\affil{2}\CorrespondingAuthor{hz2019@ic.ac.uk}
	}

\SetAffiliation{1}{Zhejiang University, Hangzhou, China}
\SetAffiliation{2}{Dyson School of Design Engineering, London, England}


\maketitle





\begin{abstract}

TRIZ, the Theory of Inventive Problem Solving, is derived from a comprehensive analysis of patents across various domains, offering a framework and practical tools for problem-solving. Despite its potential to foster innovative solutions, the complexity and abstractness of TRIZ methodology often make its acquisition and application challenging. This often requires users to have a deep understanding of the theory, as well as substantial practical experience and knowledge across various disciplines. The advent of Large Language Models (LLMs) presents an opportunity to address these challenges by leveraging their extensive knowledge bases and reasoning capabilities for innovative solution generation within TRIZ-based problem-solving process. This study explores and evaluates the application of LLMs within the TRIZ-based problem-solving process. The construction of TRIZ case collections establishes a solid empirical foundation for our experiments and offers valuable resources to the TRIZ community. A specifically designed workflow, utilizing step-by-step reasoning and evaluation-validated prompt strategies, effectively transforms concrete problems into TRIZ problems and finally generates inventive solutions. Finally, we present a case study in mechanical engineering field that highlights the practical application of this LLM-augmented method. It showcases GPT-4's ability to generate solutions that closely resonate with original solutions and suggests more implementation mechanisms. 

\end{abstract}



\section{Introduction}
TRIZ (The Theory of Inventive Problem Solving), as a design methodology derived from a large number of patents across various domains, harbors the potential for innovative solutions. Compared to conventional problem-solving methods like brainstorming, which seek specific solutions to specific problems, a key feature of TRIZ is its systematic approach of abstracting the problem to a higher level before seeking a solution \cite{ilevbareReviewTRIZIts2013}. This systematic approach serves as a guide for applying TRIZ method across various domains. Hence, although the theory was initially developed by Altshuller focusing on mechanical solutions \cite{chechurinUnderstandingTRIZReview2016}, TRIZ has been extensively promoted and applied in numerous domains \cite{orloffModernTRIZPractical2012}.

Despite the potential of the TRIZ methodology, challenges associated with the acquisition and application of TRIZ knowledge have emerged, as highlighted in a survey of TRIZ enthusiasts \cite{chechurinUnderstandingTRIZReview2016}. The complexity of TRIZ tools, such as the abstract nature of its 40 inventive principles, poses significant understanding and application difficulties \cite{cortesroblesCasebasedReasoningTRIZ2009, orloffInventiveThinkingTRIZ2006}. Regarding the essence of the problem-solving process, designers ultimately strive for effective solutions that address real-world challenges. However, this endeavor is often hampered by the practical experience necessitated during the TRIZ-based problem-solving process. 
Previous research has primarily focused on addressing this issue from methodological approaches \cite{cortesroblesCasebasedReasoningTRIZ2009, jouCREATINGINTERACTIVEWEBBASED2010, tanProcessTwoStages2007} and computer-aided support \cite{zanni-merkUseFormalOntologies2011, berdyuginaAutomaticExtractionInventive2023}, by constructing databases of patents and cases to facilitate problem-solving. Nevertheless, these strategies still require a deep understanding of TRIZ theory and the employment of cross-domain analogical reasoning to conceive possible problem-solving solutions tailored to the current problem scenario.

Within the design community, the growing interest in the power of Large Language Models (LLMs) has led to scholarly research on their application in conceptual design \cite{maConceptualDesignGeneration2023a, wangTaskDecomposedAIAidedApproach2023b}. As large pre-trained models, LLMs possess extensive knowledge repositories and reasoning capabilities, which hold substantial promise for alleviating the time burden associated with mastering the TRIZ methodology. Additionally, they can reduce the need for practical experience and cross-disciplinary knowledge. Several potential applications of LLMs within TRIZ include standardizing the workflow of TRIZ-based problem-solving process and enhancing the solution generation phase. 

Nevertheless, the problem-solving focus of TRIZ necessitates a thoughtful development and careful evaluation of the approach's reliability and efficacy. In assessing reliability, it's crucial to examine whether LLMs have accurately understood the problem context and have effectively retrieved the correct TRIZ knowledge. On the other hand, efficacy involves evaluating the reasoning capabilities of LLMs, their accuracy in applying the TRIZ systematic approach, and the effectiveness of the generated solutions. Consequently, to evaluate LLMs' abilities in problem understanding, TRIZ knowledge comprehension, and reasoning capabilities during TRIZ-based problem solving process, it is reasonable to consider the previous successful TRIZ-application cases as benchmark. This task can be challenging, due to the scarcity of comprehensive TRIZ case collections and the fact that a significant number of TRIZ projects have been abstracted the implementation details owing to confidentiality issues.

To address this, we first conducted an extensive literature review and manually curated TRIZ cases. To facilitate the evaluation of performance differences between large language models, specifically GPT-3.5 and GPT-4, we divided the cases into two collections. Collection A comprises 37 classic TRIZ cases, while Collection B includes 10 cases beyond the training dataset of GPT-4.

Subsequently, we introduced a workflow that augments TRIZ with LLMs, using step-by-step reasoning and evaluation-validated prompt engineering to augment the traditional TRIZ-based problem-solving process. To improve the transparency of the workflow's effect, we conducted quantitative evaluation experiments on four prompt strategies using case collection A. These included assessments of TRIZ contradiction analysis and solutions reasoning. And case collection B was utilized for assessing the performance differences between GPT-4 and GPT-3.5. 

Finally, leveraging the proposed workflow with validated prompt strategies, we conducted a case study in mechanical engineering domain about in-pipe robot design to showcase the method's practical application. We primarily focused on the inventive solutions generation phase, examining the solutions generated by GPT-4 and GPT-3.5 in response to TRIZ inventive principles. To visually present the semantic relationships between solutions generated by the LLMs and actual solutions, we encoded both with keywords and performed a visualization analysis. This analysis showed that solutions generated by GPT-4 align more closely with real solutions and provided a more detailed information regarding potential solution mechanisms.

\begin{figure*}[htp]
    \centering
    \includegraphics[width=0.9\linewidth]{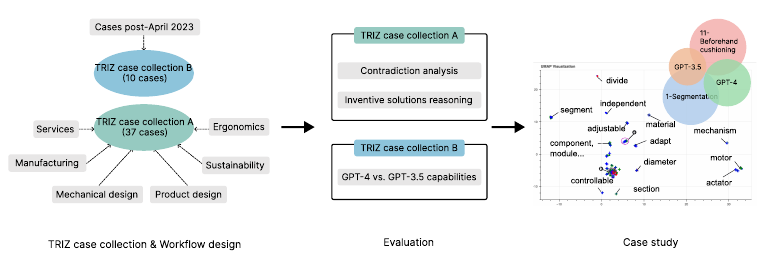}
    \caption{Overview of our contributions.}
    \label{fig: overview}
    \vspace{-0.1in}
\end{figure*}

The main contributions of our research is illustrated in Figure \ref{fig: overview}. Specifically, with TRIZ-GPT, we present the following contributions:
\begin{itemize}
  \item \textbf{TRIZ Case Collections:} We developed two TRIZ case collections categorized by the online time. Collection A compiles 37 classic TRIZ examples, while Collection B includes 10 cases published after April 2023. Considering the scarcity of datasets associated with TRIZ, these collections not only provide a solid foundation for our experiments but also serve as valuable resources for the TRIZ community.
    
  \item \textbf{Workflow Design:} Drawing from the fundamental approach of TRIZ for problem-solving, we designed an LLM-augmented workflow. For the critical stages of contradiction analysis and solution reasoning, we utilized verified prompt strategies. Through experimental validation, we applied the Chain-of-Thought and few-shot techniques to the respective stages of our proposed workflow.
    
  \item \textbf{Evaluations and Case Study:} Through evaluations to four prompt strategies, we assessed the TRIZ contradiction analysis and solution reasoning capabilities quantitatively for determining the most applicable strategy. We also conducted a comparative analysis of the performances of GPT-4 and GPT-3.5 during the contradiction analysis stage. A case study in mechanical engineering on in-pipe robot design demonstrated the practical application of our proposed method.

\end{itemize}

\section{Background}
\subsection{TRIZ for problem-solving}
TRIZ, as an innovative method for problem-solving, has been widely applied in design field, particularly in the stage of solution exploration \cite{cavallucciTRIZTheoryInventive2017a, haynesSupersystemDigitalTwindriven2023, dongProductRedesignUsing2021}. The principle of contradiction, a key concept in TRIZ, frames technical problems using parameters that either improve or worsen system conditions \cite{terninko1998systematic}. By employing the contradiction matrix, where the improving parameter and worsening parameter serve as the vertical and horizontal axes, respectively, designers can pinpoint the recommended inventive principles of TRIZ \cite{altshuller200240}. This process embodies the fundamental TRIZ paradigm: it starts with a "concrete problem", which is then translated into a "TRIZ problem". Subsequently, it evolves into a "TRIZ solution", and finally derives the "concrete solution". 


While showing application-oriented contributions across many fields, some issues have emerged in practical scenarios surrounding TRIZ. From the methodological viewpoint, the complexities and abstract nature of TRIZ make it time-consuming for many to master \cite{ilevbareReviewTRIZIts2013}. From the user's perspective, while TRIZ principles guide problem-solving across various fields, their high level of abstraction can make it difficult for designers to relate them to specific problem contexts. This is particularly acute with principles like "7-Nested Doll" or "11-Beforehand Cushioning", which require designers to independently conceive the mechanisms or materials based on the stimuli. These two factors limit the effectiveness and efficiency when using TRIZ in solving problems. Conventionally, designers start by searching the internet for related patents and cases, seeking inspiration and references for possible solutions. However, due to the vast and scattered nature of knowledge on the internet, it is usually challenging to consult the patents and papers, which makes this process time-consuming and labor-intensive \cite{casciniPlasticsDesignIntegrating2004a}. Besides, finding reference solutions closely aligned with their specific research area is often unrealistic. Thus, design teams are still required to engage in brainstorming sessions based on the information gathered \cite{chechurinResearchPracticeTheory2016}.

To alleviate the cognitive requirements of users when applying TRIZ, researchers have conducted studies in both methodological improvements and computer-aided innovation (CAI) software. In combination with other design theories, TRIZ is primarily utilized to identify inventive principles, while other design strategies offer references to aid designers in conceiving specific solutions, such as case-based reasoning (CBR) \cite{cortesroblesCasebasedReasoningTRIZ2009, jouCREATINGINTERACTIVEWEBBASED2010} and analogy-based design (ABD) \cite{tanProcessTwoStages2007}. CBR concentrates on extracting historically successful cases based on their relevance to the current problem. Although this model evolves by continuously integrating new cases into its repository, its utility is limited for designers when no similar solved cases or insufficiently similar cases exist in the case base, offering only inventive principles or common knowledge \cite{cortesroblesCasebasedReasoningTRIZ2009}. In the two-stage analogy-based design approach proposed by Tan \cite{tanProcessTwoStages2007}, it is noted that specialized methods or tools for directly transforming to domain-specific solutions are absent. The process is heavily reliant on manual effort, where designers might incorporate factors like unexpected discoveries to assist in conceptualizing solutions. Additionally, CAI software aims to automate the TRIZ-based problem-solving process utilizing artificial intelligence technology \cite{chechurinResearchPracticeTheory2016}. Designers can access case references from a semantic network driven by data, exploring cases based on semantic similarity. Research in CAI software research primarily focused on creating computerized platforms that support the application of TRIZ concepts, facilitating the use by engineering designers. For instance, the study referenced in \cite{zanni-merkUseFormalOntologies2011} developed a TRIZ process support software, TRIZacquisition, based on an ontology-driven knowledge framework. 

Moreover, since TRIZ is a design theory inspired by a thorough analysis from patents, it inspires us to review the recent tools that utilize and consolidate patent knowledge for TRIZ-based problem-solving. AI technology is increasingly being utilized to aid in extracting contradiction parameters from patents  \cite{berdyuginaAutomaticExtractionInventive2023, cascini2007computer}. During the solution generation phase, although some scholars have constructed semantic networks from patent information to support design solution ideation \cite{sarica2020technet, luo2019computer}, designers often find it challenging to connect these stimuli with the problems they face \cite{chen2024designfusion}.


Although aimed at overcoming the shortcomings of TRIZ, the aforementioned approaches primarily offer references to designers from the perspective of reference-stimuli and are not capable of directly generating solutions for specific problem scenarios \cite{zhuGenerativeTransformersDesign2022}. The process of conceiving solutions still relies on the designer's ability to draw cross-domain analogies and insights \cite{tanProcessTwoStages2007}. AutoTRIZ \cite{jiang2024autotriz}, a recent study which offers a design automation system based on TRIZ knowledge, promoted a novel approach for design ideation. Nevertheless, as a fully automated design system, the problem statement prompt is the designers' only intervention in the reasoning process. The system is designed to identify one TRIZ contradiction pair in each round, which may limit designers' solution space. Additionally, there may be inconsistencies between the system-generated contradiction analysis and the designers' expected directions, due to limited designer control during the reasoning process.

Overall, our study aims to utilize large language models to augment the TRIZ-based problem-solving process. This approach not only leverages the extensive knowledge base and reasoning capabilities of LLMs, but also struggle to provide designers with a more controllable workflow.

\subsection{Generative models in design solution generation}
Generative models have introduced new possibilities in design theory and methodology. Scholars are actively exploring the potential of generative models to enhance the design solution generation process, particularly in quickly obtaining solution stimuli. For example, Ma et al. \cite{maConceptualDesignGeneration2023a} conducted experiments with GPT-3 using base prompts, zero-shot prompts, and few-shot prompts across 12 design problems, compared the generated solutions to crowdsourced solutions, which showed improved feasibility and usefulness. However, this approach lacks structured guidance for solving design problems, and its outputs cannot overcome challenges such as lack of transparency and insufficient controllability. To mitigate these issues, some researchers have integrated traditional design models to better align LLMs with designers' thinking. For instance, Wang et al. \cite{wangTaskDecomposedAIAidedApproach2023b} proposed an FBS-based task-decomposed framework for generative design. Zhu et al. \cite{zhuGenerativeDesignIdeation2022} incorporated a design-by-analogy method into the GPT-3-based problem-solving prompt. And Chen et al. \cite{chen2024designfusion} enhanced the interpretability of the proposed design approach by integrating with classic design theories (i.e., 5W1H method, the FBS model, and Kansei Engineering). 

In this study, we select large language models for our TRIZ-based problem-solving process, as text-based forms are sufficiently used to present the initial solution. 
Given the importance of reliable assessments in this context, computational evaluation methods are increasingly valued. In the realm of studying large-scale, text-based design solutions generated by LLMs, computational evaluation methods are gaining increasing attention for their objectivity and efficiency, which could help overcome the subjectivity inherent in human evaluation. Researchers have conducted quantitative assessments from the perspectives of similarity and novelty, employing techniques such as Word Mover’s Distance (WMD) \cite{kusnerWordEmbeddingsDocumenta} and cosine vector embedding \cite{regenwetterStatisticalSimilarityRethinking2023}. Both computational assessment methods mentioned above require ground-truth references to measure the similarity between generated solutions and reference solutions. For example, in an analogy-based design evaluation inspired by biological concepts \cite{zhuBiologicallyInspiredDesign2023}, the Word Mover's Distance method was used to assess the capabilities of fine-tuned models in generating novel design concepts, with samples selected from an innovation dataset to ensure reliability.

\section{Methods}
In this section, we detail the construction, composition, and distinctions between TRIZ case collections A and B developed in this study. Then, we introduce the TRIZ-GPT workflow pipeline. The evaluation of TRIZ-GPT workflow based on the two TRIZ case collections and the demonstration through a case study will be elaborated in the following two sections.
\subsection{TRIZ case collections}
\label{sec:TRIZ case collection}
To facilitate the evaluation of performance differences between large language models, specifically GPT-3.5 and GPT-4 in this study, two distinct TRIZ case collections were compiled. Given that the training data for the GPT-4 model encompasses information available up to April 2023\footnote{\url{https://openai.com/research/gpt-4}}, TRIZ case collection A includes classic TRIZ cases published prior to this date, while TRIZ case collection B consists of cases published after April 2023, thereby reflecting the latest applications and updates in the field.
As for the case collection construction process, this study utilized Google Scholar, Springer, and Scopus as the primary literature search platforms, employing 'TRIZ', 'TRIZ cases', and 'TRIZ application' as search keywords to ensure relevance. It should be claimed that our aim is not to provide a comprehensive review of TRIZ cases, but to construct a case collection to facilitate the following evaluation process. Therefore, in the case selection phase of this study, emphasis was placed on cases that directly aligned with the requirements for validating the presented workflow in \ref{subsection: workflow}. This selection criteria led to the exclusion of certain types of cases. For example, studies that deduced solutions directly through TRIZ inventive principles without explicitly stating which parameters were improved or worsened \cite{barianiCombinedDFMATRIZ2004, ekmekci2015triz}, or those presenting solutions solely in pictorial form without textual descriptions \cite{tsaiOptimalDesignMetal2004}.


Our research team ultimately compiled 37 classic TRIZ application cases for testing the proposed workflow. Furthermore, 10 recent cases post-April 2023 were also included, reflecting the latest applications and updates. 
As highlighted in \cite{chechurinUnderstandingTRIZReview2016}, TRIZ extends beyond merely enhancing engineering solutions, with its application prevalent in various fields. Therefore, in curating our case collection, we focused on the diversity of domains such as product design \cite{orloffModernTRIZPractical2012}, engineering \cite{dengInnovationDesignSubmersible2010}, manufacturing \cite{casciniPlasticsDesignIntegrating2004}, sustainability \cite{yaoApplicationTRIZInnovative2022}, ergonomics \cite{akayCollaborativeToolSolving2008}, and service design \cite{shaoboIntegratedModeResearch2009}. This approach aims to enrich the diversity and coverage across multiple problem areas and types. Besides, we followed standardized engineering parameters’ labeling and naming guidelines as outlined by \cite{altshuller200240}, particularly for cases derived from the A-Matrix in relevant literature \cite{orloff2012modern, mi2006inventive}. For full dataset, please contact the corresponding author upon reasonable request.

\subsection{TRIZ-GPT workflow}
\label{subsection: workflow}
\begin{figure*}[htp]
    \centering
    \includegraphics[width=0.9\linewidth]{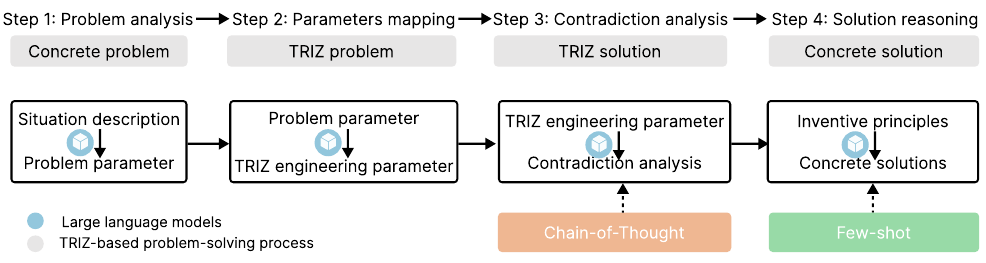}
    \caption{Illustration of the proposed TRIZ-GPT workflow.}
    \label{fig: Workflow}
     \vspace{-0.1in}
\end{figure*}

Drawing on related works and our research goals, we introduce TRIZ-GPT, an LLM-augmented TRIZ workflow designed for problem-solving. Specifically, TRIZ-GPT aims to assist designers in transforming standard problems into TRIZ problems and alleviate their cognitive load during the solution reasoning process.

Our workflow, designed based on the traditional TRIZ problem solving framework \cite{zanniFormalizationInnovatingDesign2006}, begins in Step 1 with designers inputting a problem description. Large language models would summarize the problem description into problem parameters, which represent the key factors to consider within the problem context. In Step 2, LLMs assist designers in transforming the input problem parameters into the 39 standard TRIZ engineering parameters. To enhance the standardization of output results, we integrated the engineering parameters from \cite{altshuller200240} into the LLMs and instructed them to perform parameters mapping based on the standard TRIZ parameters. It is important to note that we do not limit the output length of LLMs, fostering potential inspiration for designers. As the first two steps do not demand high reasoning capabilities of LLMs, and our preliminary tests showed that basic role positioning and task description prompts could already yield acceptable results. After reviewing the TRIZ parameters generated in Step 2, designers will select key parameters based on the problem context and input the filtered results into Step 3. In Step 3, LLMs help to analyze the user-input TRIZ parameters into contradiction pairs and elucidate their contradictions relations. As validated in preliminary experiment, this workflow is designed to assist designers in addressing one contradiction at a time for achieving more detailed output solutions. Therefore, the LLM will guide designers to select the most critical contradiction pair, providing the corresponding TRIZ inventive principles recommended by TRIZ. Finally, in Step 4, the LLM will employ these principles to reason through and generate concrete solutions based on the provided problem scenario and selected inventive principle.

To enhance the performance of LLMs in analysing TRIZ contradiction pairs and reasoning concrete solutions in Step 3 and 4, we devised four distinct prompt strategies: basic prompt, Chain-of-Thought, few-shot, and combination of Chain-of-Thought and few-shot for the two steps. The basic prompt, referred to as the "base prompt", primarily includes instructions about role positioning and task description. The other three prompts are developed further, building upon this base prompt. The Chain-of-Thought prompt features a specific reasoning path, while the few-shot prompt incorporates three examples derived from original TRIZ textbook practices. Finally, the combined prompt integrates the Chain-of-Thought's specific reasoning path with the three example scenarios used in the few-shot prompt. Due to space constraints, detailed descriptions of these four prompt strategies, as well as the prompts used in Steps 1 and 2, are provided in Appendix~\ref{appendix: prompt details}.


\section{Evaluations and Results}
\label{sec:Evaluation}
In this section, we conducted computational evaluations to determine the most suitable prompt strategy for Steps 3 and 4, and we also carried out a comparative study of GPT-4 and GPT-3.5's performance in Step 3 (the comparison for Step 4 is presented in Section \ref{sec: case study}). The evaluations in Sections \ref{subsec:step 3 eva} and \ref{subsec:step 4 eva} were based on TRIZ case collection A, while the evaluation in Section \ref{subsec: models eva} employed TRIZ case collection B.

\subsection{Evaluation of Step 3: Contradiction analysis}
\label{subsec:step 3 eva}
In this experiment, we employed the state-of-the-art model GPT-4 as the underlying large language model, setting the temperature to 0 to elicit more deterministic responses. Contradiction analysis is a fundamental step when applying TRIZ for problem-solving, laying the groundwork for determining the inventive principles for the subsequent step. In this evaluation section, we will first focus on whether the generated analysis by GPT-4 contain the contradiction pairs used in the literature. Then, we will discuss the implications of applying large language models to the contradiction analysis stage of the TRIZ methodology. 

After excluding three cases used for few-shot learning,
the remaining 34 cases were tested using a python script that called the GPT-4 model API from OpenAI. To control variables within the experiment, we ensured the [CASE\textunderscore DESCRIPTION] and [TRIZ\textunderscore PARAMETERS] remained consistent across four prompt strategy groups, varying only the prompts used to analyze the contradictory relations. Regarding the specific strategy, [TRIZ\textunderscore PARAMETERS] were derived from all transformed problem parameters in Step 2, with no parameters removed except for duplicates.

To quantitatively assess the contradiction pairs generated by GPT-4, we calculated the recall and precision scores of GPT-4's generation using the following equations, where \(\mathbf{G}_p\) denotes the contradictions generated by GPT-4, \(\mathbf{O}_p\) denotes the contradictions used in the original literature:

\[
\text{Recall} = \frac{|\mathbf{G}_p \cap \mathbf{O}_p|}{|\mathbf{O}_p|}
\]

\[
\text{Precision} = \frac{|\mathbf{G}_p \cap \mathbf{O}_p|}{|\mathbf{G}_p|}
\]
It is important to note that since there is no standardized answer for the correspondence of engineering contradictions within the problem context, the contradictions cited in the reference literature may not be comprehensive. However, this study considers these peer-reviewed analyses as acceptable answers. They provide a reliable standard for comparing the effectiveness of the four prompt strategies, aiming to avoid the potential disagreements that might arise from the subjectivity of expert assessments.

To clarify, we regard the contradiction pairs from the case collection as positive samples, which were initially annotated by one researcher and subsequently reviewed by another. Table~\ref{tab: step 3 eva results} presents the results of recall and precision scores of GPT-4’s analyses in Step 3 across four prompt strategies. Specifically, precision measures the proportion of generated data samples closely aligned with real data, while recall quantifies how comprehensively the generated parameters encompass those mentioned in the studies. Given the focus of Step 3 on evaluating the LLM's comprehension and reasoning abilities through the accuracy of contradiction relations matching, this study prioritizes recall score as the primary metric for assessing prompt strategies. As CoT group achieved the highest recall score (i.e. $R_{\text{CoT}} = 0.691$, it was selected as the prompt strategy method for enhancing LLMs' performance in contradiction analysis stage. 

\begin{figure}[htp]
    \centering
    \includegraphics[width=0.9\linewidth]{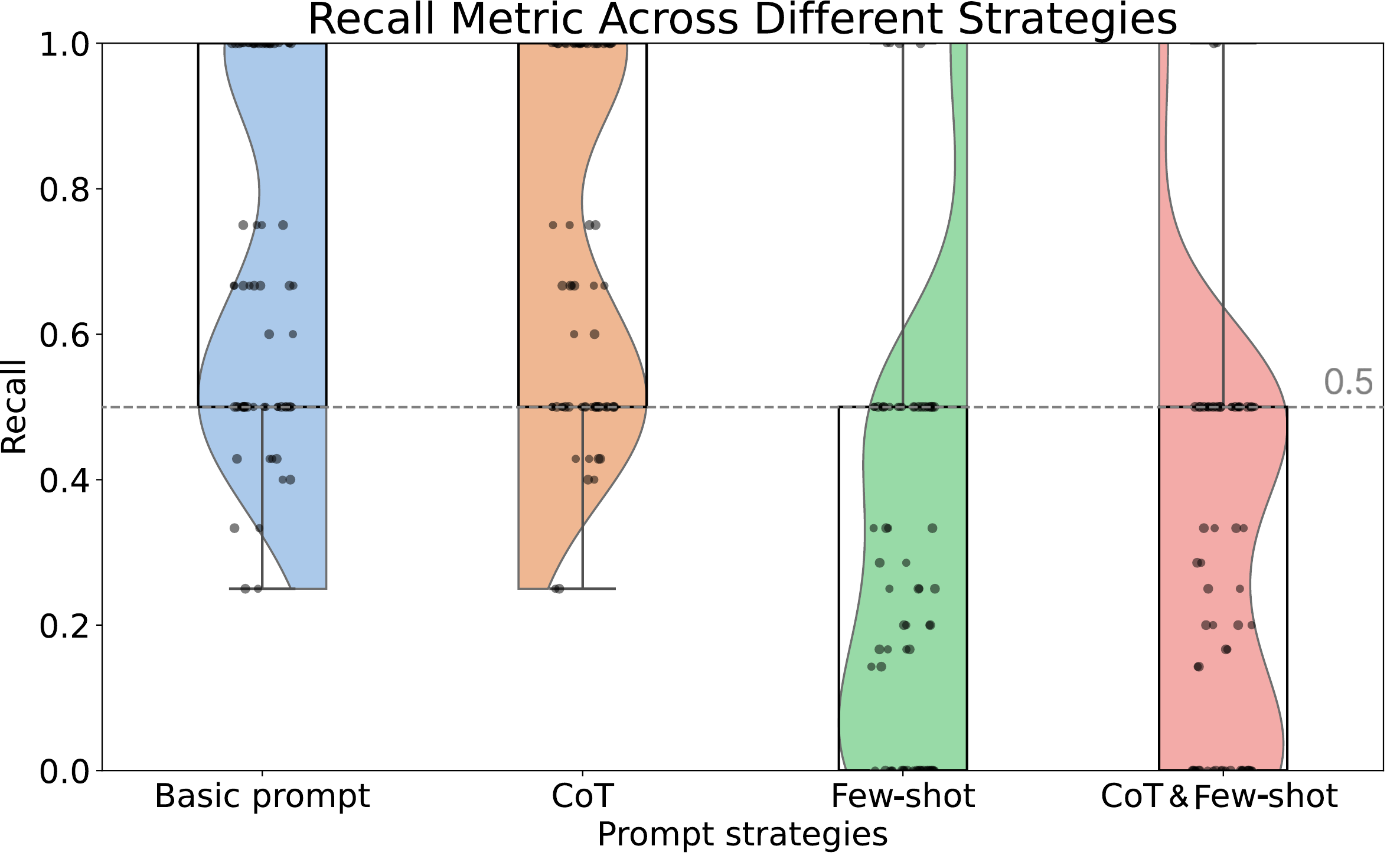}
    \caption{Recall scores across four prompt strategies during contradiction analysis.}
    \label{fig:Step 3_Precision}
    \vspace{-0.1in}
\end{figure}

Analysis of the scores from the four groups suggests that, under the condition of selecting an appropriate prompt strategy, the recall score of the data generated by LLMs is higher than the precision score in the contradiction analysis phase. This indicates that the answering generated in contradiction analysis is highly diverse. To intuitively display the distribution of score ranges across different strategies, we marked each score with a dot and plotted the violin plots, as shown in Figure \ref{fig:Step 3_Precision}. The wider sections of the violins indicate a greater number of data points, suggesting a higher concentration of recall scores within those ranges. Furthermore, manual review and calculations revealed that the majority of engineering parameters mentioned in the literature were identifiable under the conditions of either the basic prompt or the CoT prompt. Specifically, for 85.3\% of the cases using the basic prompt and 88.2\% of the cases using the CoT prompt, GPT-4 generated consistent reasoning results for at least half of the parameters in contradictory pairs. This could also be observed in Figure \ref{fig:Step 3_Precision}.

The lower recall scores observed with the few-shot strategy may stem from the LLMs unintentionally learning the number of contradiction pairs in the training samples, which led to the fewer contradiction parameters in the outputs. This is evident in Table~\ref{tab: step 3 eva results}, where the few-shot and CoT \& Few-shot groups have lower recall scores. Additionally, it can be observed from the table that the precision scores under the four prompt strategies are generally not high. This is mainly related to the data collection and precision calculation method used in this study. As previously explained at the beginning of section~\ref{subsec:step 3 eva}, all TRIZ parameters generated from the preceding step were fed into GPT-4. This may have inadvertently resulted in a reduced precision rate. However, in real-world applications, designers could analyze and filter out less feasible problem parameters or TRIZ parameters generated in Step 1 and Step 2 respectively, inputting only the most relevant ones to achieve more satisfying outcomes. Additionally, it should be noted that this experiment was designed to evaluate the effectiveness of the four prompt strategies during the contradiction analysis stage. Therefore, analyzing the problem from various contradiction views can sometimes stimulate diversity and the novelty of solutions \cite{jiang2024autotriz}. This is also consistent with our decision to use recall as the primary criterion for evaluating prompt strategies in this experiment.

\begin{table}[]
\caption{Evaluation results of four prompt strategies in contradiction analysis with GPT-4.}
\centering
\footnotesize
\label{tab: step 3 eva results}
\begin{tabular}{@{}lcccc@{}}
\toprule
          & Basic prompt     & CoT     & Few-shot     & CoT \& Few-shot     \\ \midrule
Recall    & 0.675 & 0.691 & 0.245 & 0.247 \\
Precision & 0.259 & 0.310 & 0.333 & 0.298 \\
\bottomrule
\vspace{-0.2in}

\end{tabular}
\end{table}

\subsection{Evaluation of Step 4: Solution reasoning}
\label{subsec:step 4 eva}
For this experiment, we also utilized GPT-4 as the base large language model, with the temperature parameter changed to 1 to adjust for output variability. The variation in temperature settings between the evaluations in Sections \ref{subsec:step 3 eva} and \ref{subsec:step 4 eva} is due to the distinct goals of the two steps: The contradiction analysis stage (Step 3) is designed to provide more deterministic responses regarding contradiction relationships, while the solution generation stage aims to provide a wide range of feasible design solutions, guided by inventive principles and tailored to problem scenario. 

For evaluation method, as mentioned in Section~\ref{sec:TRIZ case collection}, our case collection spans various application fields, making expert evaluations impractical. Consequently, we turned to computational evaluation metrics that are commonly used in conceptual design \cite{regenwetterStatisticalSimilarityRethinking2023}. In line with the evaluation methods selected in other conceptual design studies, such as those by \cite{maConceptualDesignGeneration2023a} and \cite{zhuGenerativeTransformersDesign2022}, and the objectives of this research, we ultimately chose to use cosine similarity to examine the similarity between the real solutions and the generated solutions. Initially, during the data collection stage, we instructed the GPT-4 model to generate three solutions for each inventive principle to enhance the reliability of the cosine similarity evaluation. Although designers in practical scenarios are free to input their selected inventive principles, for the purposes of our evaluation, we specifically used the inventive principles described in the original literature as inputs to GPT-4, ensuring a more valid comparison with the ground truth. In instances where multiple inventive principles were proposed for resolving an issue, we generated separate solutions for each principle. The total number of cases assessed in this experiment amounted to 84.

\begin{figure}[htp]
    \centering
    \includegraphics[width=0.9\linewidth]{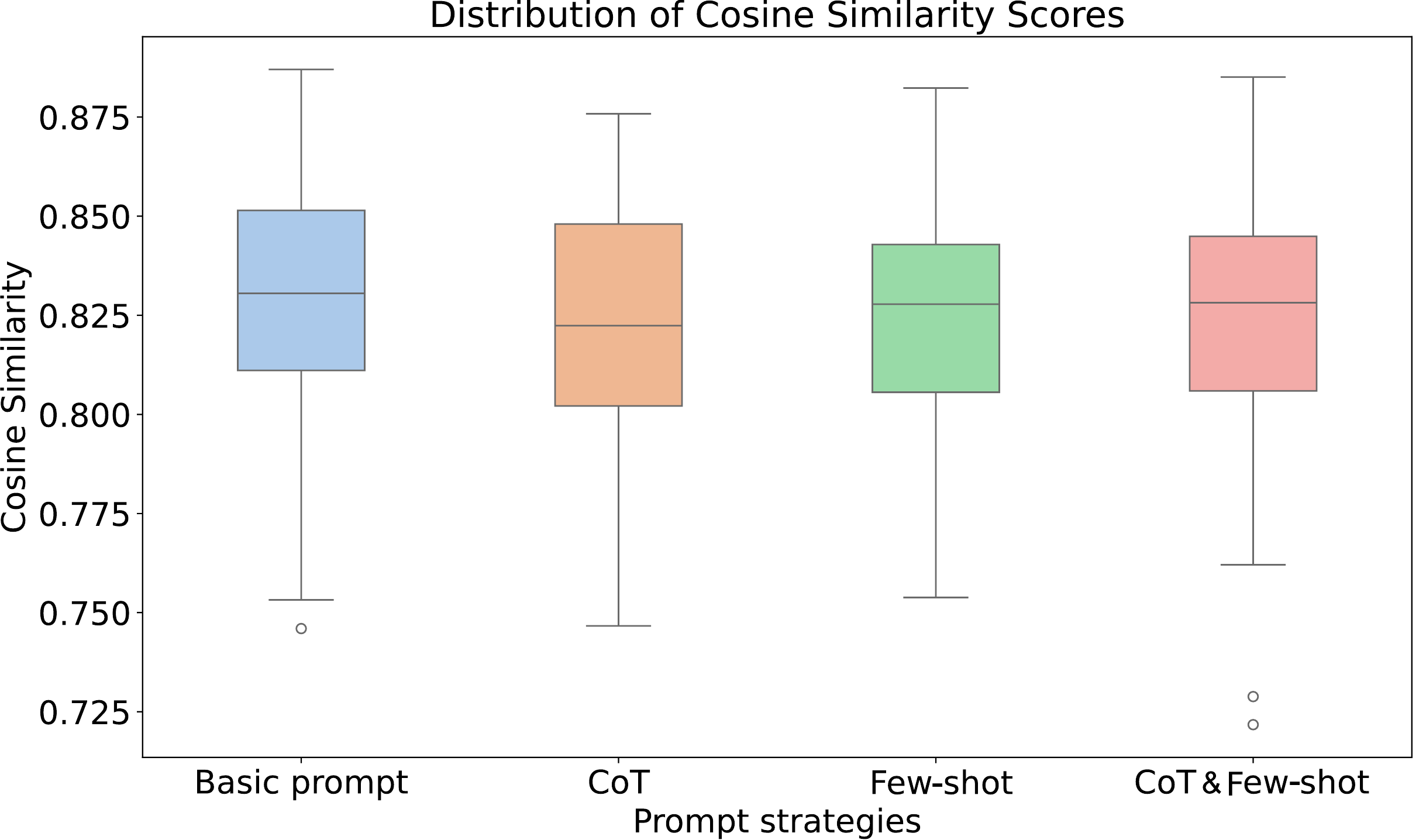}
    \caption{Distribution of the cosine similarity of four prompt strategies for solution reasoning with GPT-4}
    \label{fig: cosine similarity}
    \vspace{-0.1in}
\end{figure}

During the data analysis phase, we initially extracted solution descriptions from the original literature to serve as the ground truth. Subsequently, we computed the cosine similarity for each of the three solutions against the original solution, averaging the scores to assess GPT-4's performance in each case. This computation was carried out using OpenAI’s text-embedding-ada-002 model, and the results are presented in Figure \ref{fig: cosine similarity}.


As illustrated in Figure \ref{fig: cosine similarity}, the distribution and mean results of cosine similarity scores across the 84 cases did not show significant differences among the four prompt strategies. The GPT-4 model consistently achieves high cosine similarity with the original solutions for all strategies, with scores all exceeding 0.82 (Base prompt: \textit{M} = 0.829, \textit{SD} = 0.031; CoT prompt: \textit{M} = 0.823, \textit{SD} = 0.031; Few-shot prompt: \textit{M} = 0.824, \textit{SD} = 0.030; CoT+ Few-shot prompt: \textit{M} = 0.824, \textit{SD} = 0.033). Based on the above evaluation results and our manual review of the generated solutions, we selected the few-shot method as the prompt strategy for Step 4. This decision was driven by the few-shot strategy's ability to yield structured and stable outcomes, and its superior efficiency in utilizing input tokens to achieve comparable cosine similarity scores, relative to the combination of Chain-of-Thought and few-shot. This efficiency is crucial for maintaining clarity and conciseness.

Combining the evaluation results from Step 3 and Step 4, it is noteworthy that the combination of Chain-of-Thought and few-shot did not manifest in a clear-cut additive or synergistic impact on performance. These findings are consistent with those reported by other similar studies about the evaluation of different prompt strategies \cite{singhalLargeLanguageModels2023}. During the contradiction analysis phase in Step 3, the diversity of LLMs' outputs might have been constrained by the number of contradiction pairs given in the learning samples. In the solution reasoning phase of Step 4, base prompt may already enable GPT-4 to leverage its knowledge base and reasoning capabilities. This might be attributed to the precise role positioning prompts aiding LLMs in accessing TRIZ-related knowledge and task description prompts clarifying the output content. At this point, learning shots assist in stabilizing the output structure. Nonetheless, specifying a single reasoning path with CoT does not show significant improvement.

\subsection{Evaluation of different models}
\label{subsec: models eva}
To assess reasoning capabilities between GPT-3.5 and GPT-4, we utilized TRIZ case collection B, comprising ten recent cases published online post-April 2023. This selection postdates GPT-4's training data, thus negating potential bias from prior learning. These ten cases primarily originate from recent academic conferences on TRIZ, ensuring the timeliness of the research questions. 

The comparative experiment in this section was conducted following the same procedure in Section \ref{subsec:step 3 eva}. Using python scripts to invoke OpenAI's GPT-4 and GPT-3.5-turbo API. Given the calculation results about cosine similarity in \ref{subsec:step 4 eva}, we opted to test the performance of all cases within the TRIZ case collection B during the contradiction analysis evaluation. For the inventive solution reasoning phase, we will focus on a detailed discussion surrounding one certain case, with the specific process showcased in Section \ref{sec: case study}. 

\begin{figure}[htp]
    \centering
    \includegraphics[width=0.9\linewidth]{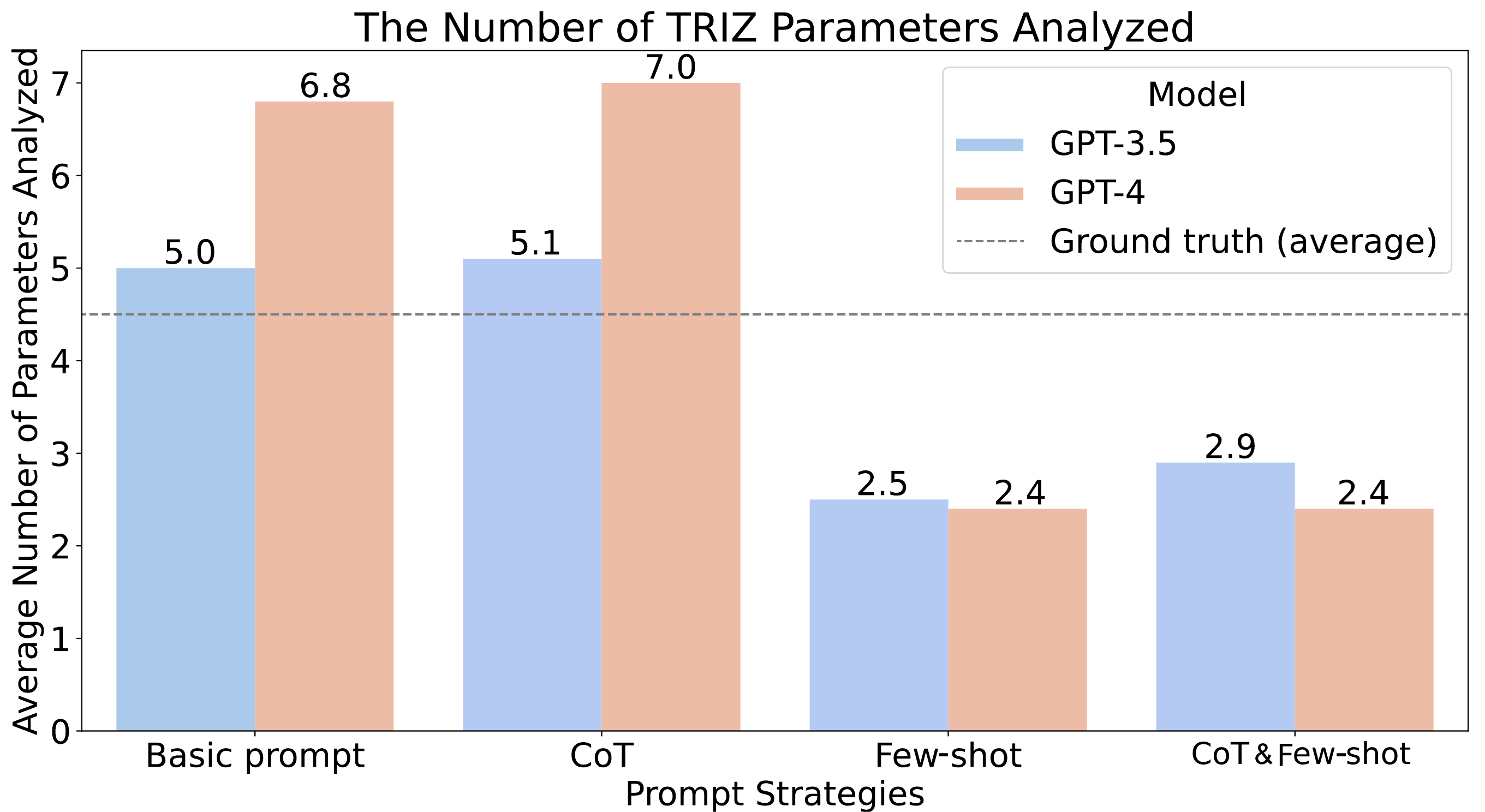}
    \caption{Comparison of the averaged number of TRIZ engineering parameters analysed by GPT-3.5 and GPT-4 under different prompt strategies.}
    \label{fig:GPTs_生成参数个数对比}
    \vspace{-0.1in}
\end{figure}

\begin{table}[htp]
\centering
\vspace{-0.04in}
\caption{Evaluation results of four prompt strategies in contradiction analysis with GPT-4 and GPT-3.5.}
\label{tab:models test_step3}
\small
 \begin{tabular}{lcccc}
  \toprule
  & \multicolumn{2}{c}{\textbf{GPT-3.5}} & \multicolumn{2}{c}{\textbf{GPT-4}} \\
  \midrule
  \textbf{ } & \textbf{recall} & \textbf{precision} & \textbf{recall} & \textbf{precision}\\
  \midrule

Basic prompt & 0.400 & 0.357 & 0.544 & 0.331\\
CoT & 0.400 & 0.328 & 0.544 & 0.327\\
Few-shot & 0.304 & 0.483 & 0.288 & 0.400\\
CoT \& Few-shot & 0.383 & 0.513 & 0.275 & 0.3833\\

  \bottomrule
 \end{tabular}
\vspace{-0.04in}
\end{table}

The results are presented in Table \ref{tab:models test_step3}. In general, both GPT models were capable of inferring the contradiction pairs used in the literature. Notably, GPT-4, employing CoT or basic prompt, achieved the highest recall scores. During our manual review process of the experiment results, we noted a significant difference in the number of contradiction pairs inferred by the two models. This observation inspired us to quantify the number of TRIZ engineering parameters analysed by each model across the four prompt strategies. The comparative results are illustrated in Figure \ref{fig:GPTs_生成参数个数对比}.

The ten cases in TRIZ case collection B involved an average of 5.5 contradiction pairs each, whereas both models under basic prompt and CoT strategies inferred more contradiction pairs than this average. Furthermore, GPT-4 was able to reason out a greater number of contradiction pairs compared to GPT-3.5, offering designers more insights for exploring TRIZ-based solutions. This capacity to generate additional contradiction pairs is one reason why GPT-4 achieved higher recall score than GPT-3.5 under the same CoT reasoning path (0.544 and 0.400 for GPT-4 and GPT-3.5 respectively). To broaden the scope of reasoning when applying GPT-3.5, it is recommended that designers guide GPT-3.5 model to expand the mapping of contradiction pairs during this phase. Moreover, despite our prompts requesting the LLMs to answer with TRIZ engineering parameter numbers, responses from GPT-3.5 occasionally lacked these details, unlike those from GPT-4, which consistently incorporated them. For a design method like TRIZ, which relies on standardized processes, such inconsistencies might be problematic.


In summary, the distinctions between GPT-3.5 and GPT-4 become more apparent as the complexity of the task increases. According to OpenAI's evaluation\footnote{\url{https://openai.com/research/gpt-4}}, GPT-4 demonstrates enhanced reliability, creativity, and capability to handle nuanced instructions compared to GPT-3.5. Specifically, when dealing with tasks requiring the inference of contradiction pairs, GPT-4, employing the Chain-of-Thought strategy, achieves higher recall scores, showing greater alignment with original TRIZ practices. Its creativity is evident as it infers a greater number of contradiction pair counts, benefiting designers in solution space exploration. Furthermore, GPT-4 demonstrates superior performance in executing complex prompts. Notably, despite the prompt explicitly requesting "Remember to tell me the number of the corresponding TRIZ engineering parameters.", GPT-3.5 fails to provide the parameter numbers in 2 out of 10 instances, while GPT-4 consistently meets this requirement. Detailed examples illustrating these three points are provided in Appendix \ref{appendix: comparison of GPT-4 & GPT-3.5}.



\section{Case study}
\label{sec: case study}
In order to provide a deeper insight into the quality of solutions generated by large language models in Step 4 (Solution reasoning), and to compare the differences between solutions produced by GPT-3.5 and GPT-4, we carried out this case study about in-pipe robot design by Xie and Liu \cite{xieApplicationTRIZInnovation2023} as an example. We selected this case because of its recency and typicality: The paper was published after GPT-4, which assures it is not included in GPT-4's training data. What's more, mechanical field has traditionally been the primary domain for TRIZ applications \cite{ilevbareReviewTRIZIts2013}. Drawing on the evaluation results of prompt strategies from Section \ref{subsec:step 4 eva}, few-shot strategy was selected as prompt engineering method to derive solutions for GPT-4 and GPT-3.5 in this experiment. 

\begin{figure*}[h]
    \centering
    \includegraphics[width=0.9\linewidth]{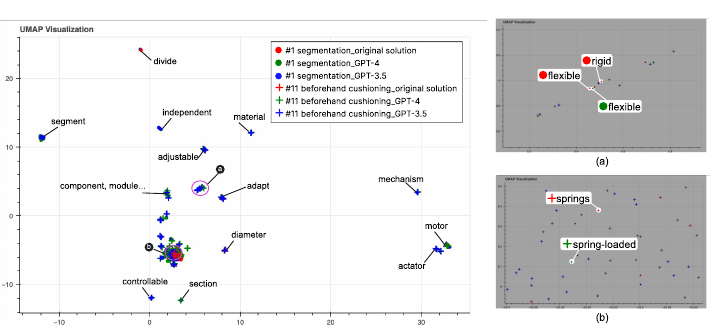}
    \caption{Visualization of keywords extracted from the original solutions and generated solutions by GPT-4 and GPT-3.5 in case study.}
    \label{fig:visualization}
     \vspace{-0.1in}
\end{figure*}

In this case, the primary technical requirements include reducing the number of motors and making the robot adapt to different pipe diameters. The application of TRIZ principles "1-Segmentation" and "11-Beforehand Cushioning" guided designers to consider solutions from two different directions. Under principle-1, designers are encouraged to think about what can be segmented and the potential outcomes of such segmentation. With principle-11, the focus shifts to identifying suitable materials or mechanisms for achieving a cushioning effect in advance.

Specifically, we directed both GPT-4 and GPT-3.5 to generate solutions one at a time, with each model producing solutions based on two distinct inventive principles. This process was repeated ten times, yielding a total of ten solutions for each principle from both language models. Upon reviewing the generated solutions, researchers found that both GPT-4 and GPT-3.5 tend to include repetitive descriptions of the problem in their outputs. While this repetition could be seen as the LLMs' attempt to echo the input text, such redundancy can complicate the analysis of the quality of the generated solutions. Therefore, we manually extracted the keywords directly related to the problem-solving mechanisms from both the original and generated solutions, while discarding irrelevant content such as design objectives or benefits. Examples of solutions generated by the two models and coding examples can be found in Appendix \ref{appendix: generated solutions and coding examples}. 

To offer a more intuitive display of the semantic distribution of the generated solutions, we computed the embeddings of the extracted keywords with GoogleNews-vectors-negative300. Subsequently, we applied Uniform Manifold Approximation and Projection (UMAP) for dimensionality reduction and visualization \cite{hamalainenEvaluatingLargeLanguage2023}. The method allows efficient visualization of the embedding vectors, as shown in 
Figure \ref{fig:visualization}. In this visualization, red markers represent the keywords of the solution used in the original literature \cite{xieApplicationTRIZInnovation2023}. Green and blue markers denote exacted keywords from GPT-4 and GPT-3.5 respectively. 



In Figure~\ref{fig:visualization}, labels such as "segment" and "divide" correlate with principle "1-Segmentation", whereas solutions pertaining to "11-Beforehand Cushioning" predominantly cluster in area (b). This visually demonstrates how different TRIZ inventive principles could guide designers to think about solving inventive problems from various directions. Additionally, it is evident that compared to the original solution represented by the red marker, the solutions generated by LLMs explored a broader semantic space. Furthermore, as illustrated in the magnified view (a) and (b) in Figure \ref{fig:visualization}, GPT-4's responses corresponded well with the keyword "flexible" for "1-Segmentation", and "spring-loaded" for "11-Beforehand Cushioning", while GPT-3.5's responses did not echo the keywords from the original solution.

\section{Discussion}
In this section, beyond synthesizing conclusions drawn from the evaluation and case study section, we delve into other insights observed during the review of LLMs' generation.
\subsection{Expanding design horizons with LLM-generated content}
Benefiting from the comprehensive nature of content generated by LLMs and the use of appropriate prompt strategies, Step 3 (Contradiction analysis) offers designers a broader space of directions for contemplation through the inferred contradiction pairs. In Step 4 (Solution reasoning), iterative generation aids in exploring potential solutions under the guidance of inventive principles, enriching designers' understanding of these principles' benefits. For instance, the case study on 1-segmentation not only highlighted known advantages like reduced manufacturing costs but also introduced new insights such as modular redundancy and ease of replacement, enhancing problem comprehension and facilitating solution development.
\subsection{LLM-augmented problem-solving process}
To evaluate the scalability issue of our proposed TRIZ-GPT methodology, which is primarily designed for the conceptual design process, we conduct a quick trial in product design. To ensure the timeliness of the case, we selected a recent product design example \cite{changTRIZinspiredKnowledgedrivenApproach2023}, focusing on the hardware design of an intelligent test tube rack. The experiment was conducted according to the conditions and process described in \ref{subsec:step 4 eva}, with results displayed in Appendix \ref{Appendix: sec_different domains}. The findings indicate that leveraging the extensive database and generative capabilities of large language models, this approach not only suggests new directions in shape design as the original solution, but also viable options for material application and color coding. These generated solutions provide practical approaches for design team to select for detailed consideration. Given the proven value and effectiveness of TRIZ theory across various domains, this quick trial also demonstrates the method's application potential to other fields. More exploration in different domains could be done such as user experience design \cite{gilani2023improving} and manufacturing design \cite{kim2022conceptual}.


\section{Limitations and future directions}
Firstly, the issue of proprietary data poses a significant challenge in the practical application of large language models. As noted by Ilevbare et al. \cite{ilevbareReviewTRIZIts2013}, detailed cases of how organizations have successfully or unsuccessfully applied TRIZ are seldom shared. As an initial step, we compiled a total of 47 TRIZ cases to validate our proposed methodology. However, future research will necessitate a more extensive collection of cases to robustly support and expand upon our findings.

Secondly, concerning the integration of TRIZ theory with large language models, our findings during the evaluation process suggest that future research could focus on enhancing the output of LLMs when combining multiple inventive principles to propose integrated solutions effectively. Additionally, investigating the use of other AI agents to assist designers in selecting more feasible inventive principles or solutions would be a valuable direction.

Thirdly, regarding the application of LLMs to empower design theories and methodologies, To ensure a reliable application of LLM-augmented design methodologies, it is essential to analyze potential biases, such as overfitting and generalization, that may arise during the training process of LLMs. We mainly discuss about the scalability issues due to space limitation and a thorough reflection on this matter requires the collaboration of the entire research community.

\section{Conclusion}

We introduce and assess TRIZ-GPT, an LLM-augmented methodology designed to facilitate problem-solving process. Leveraging the fundamental paradigms of TRIZ and the extensive knowledge base and reasoning abilities of large language models, we aim to mitigate the cognitive load and memory demands traditionally associated with the application of TRIZ. Constructing two TRIZ case collections is a time-consuming yet worthwhile endeavor. By referencing classical TRIZ practices in case collection A, we evaluated the prompt strategies for contradiction analysis and solution reasoning. Our study also examined the performance disparities between GPT-4 and GPT-3.5 in these two phases. To preclude biases due to differences in their training databases, we utilized case collection B, which lies outside of GPT-4's training repository, for assessing model capability variations. In case study, we focus on the mechanical engineering problem of in-pipe robots. Through vector dimensionality reduction and visualization, we further demonstrate that solutions generated by GPT-4 more closely align with the original solution and encompass a broader range of problem-solving strategies. Notably, for relatively abstract inventive principles, large language models can prompt designers on potential implementation mechanisms.

\section{acknowledgement}
We thank four reviewers' valuable comments to help us improve our work. This research was funded by National Key R\&D Program of China (2022YFB3303301).




\bibliographystyle{asmeconf}  
\bibliography{0305}

\begin{thebibliography}{10}
\newcommand{\enquote}[1]{``#1''}
\providecommand{\url}[1]{\texttt{#1}}
\providecommand{\urlprefix}{URL }
\expandafter\ifx\csname urlstyle\endcsname\relax
  \providecommand{\doi}[1]{DOI \discretionary{}{}{}#1}\else
  \providecommand{\doi}{DOI \discretionary{}{}{}\begingroup
  \urlstyle{rm}\Url}\fi
\providecommand{\eprint}[2][]{\urlprefix\href{#1#2}{#1#2}}

\bibitem{ilevbareReviewTRIZIts2013}
Ilevbare, Imoh~M., Probert, David and Phaal, Robert.
\newblock \enquote{A Review of {{TRIZ}}, and Its Benefits and Challenges in
  Practice.}
\newblock \textit{Technovation} Vol.~33 No. 2-3 (2013): pp. 30--37.
\newblock \doi{10.1016/j.technovation.2012.11.003}.

\bibitem{chechurinUnderstandingTRIZReview2016}
Chechurin, Leonid and Borgianni, Yuri.
\newblock \enquote{Understanding {{TRIZ}} through the Review of Top Cited
  Publications.}
\newblock \textit{Computers in Industry} Vol.~82 (2016): pp. 119--134.
\newblock \doi{10.1016/j.compind.2016.06.002}.

\bibitem{orloffModernTRIZPractical2012}
Orloff, Michael~A.
\newblock \textit{Modern {{TRIZ}}: {{A Practical Course}} with {{EASyTRIZ
  Technology}}}.
\newblock {Springer Berlin Heidelberg}, {Berlin, Heidelberg} (2012).
\newblock \doi{10.1007/978-3-642-25218-1}.

\bibitem{cortesroblesCasebasedReasoningTRIZ2009}
Cortes~Robles, Guillermo, Negny, St{\'e}phane and Le~Lann, Jean~Marc.
\newblock \enquote{Case-Based Reasoning and {{TRIZ}}: {{A}} Coupling for
  Innovative Conception in {{Chemical Engineering}}.}
\newblock \textit{Chemical Engineering and Processing: Process Intensification}
  Vol.~48 No.~1 (2009): pp. 239--249.
\newblock \doi{10.1016/j.cep.2008.03.016}.

\bibitem{orloffInventiveThinkingTRIZ2006}
Orloff, Michael~A.
\newblock \textit{Inventive Thinking through {{TRIZ}}: A Practical Guide}, 2nd
  ed.
\newblock {Springer}, {Berlin ; New York} (2006).

\bibitem{jouCREATINGINTERACTIVEWEBBASED2010}
Jou, Min and Chuang, Chien-Pen.
\newblock \enquote{{{CREATING INTERACTIVE WEB-BASED ENVIRONMENTS TO SCAFFOLD
  CREATIVE REASONING AND MEANINGFUL LEARNING}}: {{FROM PHYSICS TO PRODUCTS}}.}
\newblock \textit{The Turkish Online Journal of Educational Technology} Vol.~9
  No.~4.

\bibitem{tanProcessTwoStages2007}
Tan, Runhua.
\newblock \enquote{Process of Two Stages {{Analogy-based Design}} Employing
  {{TRIZ}}.}
\newblock \textit{International Journal of Product Development} Vol.~4 No. 1/2
  (2007): p. 109.
\newblock \doi{10.1504/IJPD.2007.011537}.

\bibitem{zanni-merkUseFormalOntologies2011}
{Zanni-Merk}, Cecilia, Cavallucci, Denis and Rousselot, Fran{\c c}ois.
\newblock \enquote{Use of Formal Ontologies as a Foundation for Inventive
  Design Studies.}
\newblock \textit{Computers in Industry} Vol.~62 No.~3 (2011): pp. 323--336.
\newblock \doi{10.1016/j.compind.2010.09.007}.

\bibitem{berdyuginaAutomaticExtractionInventive2023}
Berdyugina, Daria and Cavallucci, Denis.
\newblock \enquote{Automatic Extraction of Inventive Information out of Patent
  Texts in Support of Manufacturing Design Studies Using {{Natural Languages
  Processing}}.}
\newblock \textit{Journal of Intelligent Manufacturing} Vol.~34 No.~5 (2023):
  pp. 2495--2509.
\newblock \doi{10.1007/s10845-022-01943-y}.

\bibitem{maConceptualDesignGeneration2023a}
Ma, Kevin, Grandi, Daniele, McComb, Christopher and {Goucher-Lambert}, Kosa.
\newblock \enquote{Conceptual {{Design Generation Using Large Language
  Models}}.} .

\bibitem{wangTaskDecomposedAIAidedApproach2023b}
Wang, Boheng, Zuo, Haoyu, Cai, Zebin, Yin, Yuan, Childs, Peter, Sun, Lingyun
  and Chen, Liuqing.
\newblock \enquote{A {{Task-Decomposed AI-Aided Approach}} for {{Generative
  Conceptual Design}}.} .

\bibitem{cavallucciTRIZTheoryInventive2017a}
Cavallucci, Denis (ed.).
\newblock \textit{{{TRIZ}} -- {{The Theory}} of {{Inventive Problem Solving}}}.
\newblock {Springer International Publishing}, {Cham} (2017).
\newblock \doi{10.1007/978-3-319-56593-4}.

\bibitem{haynesSupersystemDigitalTwindriven2023}
Haynes, Patrick and Yang, Sheng.
\newblock \enquote{Supersystem Digital Twin-Driven Framework for New Product
  Conceptual Design.}
\newblock \textit{Advanced Engineering Informatics} Vol.~58 (2023): p. 102149.
\newblock \doi{10.1016/j.aei.2023.102149}.

\bibitem{dongProductRedesignUsing2021}
Dong, Yafan, Tan, Runhua, Zhang, Peng, Peng, Qingjin and Shao, Peng.
\newblock \enquote{Product Redesign Using Functional Backtrack with Digital
  Twin.}
\newblock \textit{Advanced Engineering Informatics} Vol.~49 (2021): p. 101361.
\newblock \doi{10.1016/j.aei.2021.101361}.

\bibitem{terninko1998systematic}
Terninko, John, Zusman, Alla and Zlotin, Boris.
\newblock \textit{Systematic innovation: an introduction to TRIZ (theory of
  inventive problem solving)}.
\newblock CRC press (1998).

\bibitem{altshuller200240}
Altshuller, Genrich.
\newblock \textit{40 principles: TRIZ keys to technical innovation}.
\newblock Vol.~1.
\newblock Technical Innovation Center, Inc. (2002).

\bibitem{casciniPlasticsDesignIntegrating2004a}
Cascini, Gaetano and Rissone, Paolo.
\newblock \enquote{Plastics Design: Integrating {{TRIZ}} Creativity and
  Semantic Knowledge Portals.}
\newblock \textit{Journal of Engineering Design} Vol.~15 No.~4 (2004): pp.
  405--424.
\newblock \doi{10.1080/09544820410001697208}.

\bibitem{chechurinResearchPracticeTheory2016}
Chechurin, Leonid (ed.).
\newblock \textit{Research and {{Practice}} on the {{Theory}} of {{Inventive
  Problem Solving}} ({{TRIZ}})}.
\newblock {Springer International Publishing}, {Cham} (2016).
\newblock \doi{10.1007/978-3-319-31782-3}.

\bibitem{cascini2007computer}
Cascini, Gaetano and Russo, Davide.
\newblock \enquote{Computer-aided analysis of patents and search for TRIZ
  contradictions.}
\newblock \textit{International Journal of Product Development} Vol.~4 No. 1-2
  (2007): pp. 52--67.

\bibitem{sarica2020technet}
Sarica, Serhad, Luo, Jianxi and Wood, Kristin~L.
\newblock \enquote{TechNet: Technology semantic network based on patent data.}
\newblock \textit{Expert Systems with Applications} Vol. 142 (2020): p. 112995.

\bibitem{luo2019computer}
Luo, Jianxi, Sarica, Serhad and Wood, Kristin~L.
\newblock \enquote{Computer-aided design ideation using InnoGPS.}
\newblock \textit{International design engineering technical conferences and
  computers and information in engineering conference}, Vol. 59186: p.
  V02AT03A011. 2019. American Society of Mechanical Engineers.

\bibitem{chen2024designfusion}
Chen, Liuqing, Jing, Qianzhi, Tsang, Yixin, Wang, Qianyi, Sun, Lingyun and Luo,
  Jianxi.
\newblock \enquote{DesignFusion: Integrating Generative Models for Conceptual
  Design Enrichment.}
\newblock \textit{Journal of Mechanical Design}  (2024): pp. 1--61.

\bibitem{zhuGenerativeTransformersDesign2022}
Zhu, Qihao and Luo, Jianxi.
\newblock \enquote{Generative {{Transformers}} for {{Design Concept
  Generation}}.}
\newblock \textit{Journal of Computing and Information Science in Engineering}
  (2022): pp. 1--61\doi{10.1115/1.4056220}.

\bibitem{jiang2024autotriz}
Jiang, Shuo and Luo, Jianxi.
\newblock \enquote{AutoTRIZ: Artificial Ideation with TRIZ and Large Language
  Models.} (2024).
\newblock \eprint{2403.13002}.

\bibitem{zhuGenerativeDesignIdeation2022}
Zhu, Qihao and Luo, Jianxi.
\newblock \enquote{Generative {{Design Ideation}}: {{A Natural Language
  Generation Approach}}.} (2022).
\newblock \eprint{2204.09658}.

\bibitem{kusnerWordEmbeddingsDocumenta}
Kusner, Matt~J, Sun, Yu, Kolkin, Nicholas~I and Weinberger, Kilian~Q.
\newblock \enquote{From {{Word Embeddings To Document Distances}}.} .

\bibitem{regenwetterStatisticalSimilarityRethinking2023}
Regenwetter, Lyle, Srivastava, Akash, Gutfreund, Dan and Ahmed, Faez.
\newblock \enquote{Beyond {{Statistical Similarity}}: {{Rethinking Metrics}}
  for {{Deep Generative Models}} in {{Engineering Design}}.} (2023).
\newblock \eprint{2302.02913}.

\bibitem{zhuBiologicallyInspiredDesign2023}
Zhu, Qihao, Zhang, Xinyu and Luo, Jianxi.
\newblock \enquote{Biologically {{Inspired Design Concept Generation Using
  Generative Pre-Trained Transformers}}.}
\newblock \textit{Journal of Mechanical Design} Vol. 145 No. 041409.
\newblock \doi{10.1115/1.4056598}.

\bibitem{barianiCombinedDFMATRIZ2004}
Bariani, P.~F., Berti, G.~A. and Lucchetta, G.
\newblock \enquote{A {{Combined DFMA}} and {{TRIZ}} Approach to the
  Simplification of Product Structure.}
\newblock \textit{Proceedings of the Institution of Mechanical Engineers, Part
  B: Journal of Engineering Manufacture} Vol. 218 No.~8 (2004): pp. 1023--1027.
\newblock \doi{10.1243/0954405041486091}.

\bibitem{ekmekci2015triz}
Ekmekci, Ismail and Koksal, Mustafa.
\newblock \enquote{Triz methodology and an application example for product
  development.}
\newblock \textit{Procedia-Social and Behavioral Sciences} Vol. 195 (2015): pp.
  2689--2698.

\bibitem{tsaiOptimalDesignMetal2004}
Tsai, C.C., Chang, C.Y. and Tseng, C.H.
\newblock \enquote{Optimal Design of Metal Seated Ball Valve Mechanism.}
\newblock \textit{Structural and Multidisciplinary Optimization} Vol.~26 No.
  3-4 (2004): pp. 249--255.
\newblock \doi{10.1007/s00158-003-0342-3}.

\bibitem{dengInnovationDesignSubmersible2010}
Deng, Jun, Feng, Li-jie and Wang, Jin-feng.
\newblock \enquote{Innovation Design of Submersible Motor Pump Based on
  {{Triz}}.}
\newblock \textit{2010 {{IEEE 17Th International Conference}} on {{Industrial
  Engineering}} and {{Engineering Management}}}: pp. 520--523. 2010. {IEEE},
  {Xiamen, China}.
\newblock \doi{10.1109/ICIEEM.2010.5646562}.

\bibitem{casciniPlasticsDesignIntegrating2004}
Cascini, Gaetano and Rissone, Paolo.
\newblock \enquote{Plastics Design: Integrating {{TRIZ}} Creativity and
  Semantic Knowledge Portals.}
\newblock \textit{Journal of Engineering Design} Vol.~15 No.~4 (2004): pp.
  405--424.
\newblock \doi{10.1080/09544820410001697208}.

\bibitem{yaoApplicationTRIZInnovative2022}
Yao, Kai-Chao, Li, Kuo-Yi, Xu, Jing-Ran, Ho, Wei-Sho and Shen, Yu-Hao.
\newblock \enquote{Application of {{TRIZ Innovative System Method}} in {{Rapid
  Assembly}} of {{Folding Chairs}}.}
\newblock \textit{Sustainability} Vol.~14 No.~22 (2022): p. 15482.
\newblock \doi{10.3390/su142215482}.

\bibitem{akayCollaborativeToolSolving2008}
Akay, D., Dem{\i}ray, A. and Kurt, M.
\newblock \enquote{Collaborative Tool for Solving Human Factors Problems in the
  Manufacturing Environment: The {{Theory}} of {{Inventive Problem Solving
  Technique}} ({{TRIZ}}) Method.}
\newblock \textit{International Journal of Production Research} Vol.~46 No.~11
  (2008): pp. 2913--2925.
\newblock \doi{10.1080/00207540600969774}.

\bibitem{shaoboIntegratedModeResearch2009}
Shaobo, Li, Yuqin, Ma, Guanci, Yang and Yaqing, Li.
\newblock \enquote{An {{Integrated Mode Research}} of {{QFD}} and {{TRIZ}} and
  {{Its Applications}}.}
\newblock \textit{2009 {{Second International Workshop}} on {{Computer
  Science}} and {{Engineering}}}: pp. 548--552. 2009. {IEEE}, {Qingdao, China}.
\newblock \doi{10.1109/WCSE.2009.729}.

\bibitem{orloff2012modern}
Orloff, Michael~A.
\newblock \textit{Modern TRIZ: A practical course with easytriz technology}.
\newblock Springer Science \& Business Media (2012).

\bibitem{mi2006inventive}
Mi~Dahlgaard-Park, Su.
\newblock \enquote{Inventive thinking through TRIZ: a practical guide.}
\newblock \textit{The TQM Magazine} Vol.~18 No.~3 (2006): pp. 312--314.

\bibitem{zanniFormalizationInnovatingDesign2006}
Zanni, Cecilia and Rousselot, Fran{\c c}ois.
\newblock \enquote{Towards the {{Formalization}} of {{Innovating Design}}:
  {{The TRIZ Example}}.}
\newblock Hutchison, David, Kanade, Takeo, Kittler, Josef, Kleinberg, Jon~M.
  and Mattern, Friedemann et~al. (eds.). \textit{Knowledge-{{Based Intelligent
  Information}} and {{Engineering Systems}}}.
\newblock Vol. 4251. {Springer Berlin Heidelberg}, {Berlin, Heidelberg} (2006):
  pp. 1098--1105.
\newblock \doi{10.1007/11892960_132}.

\bibitem{singhalLargeLanguageModels2023}
Singhal, Karan, Azizi, Shekoofeh, Tu, Tao, Mahdavi, S.~Sara, Wei, Jason, Chung,
  Hyung~Won, Scales, Nathan and Tanwani, et~al., Ajay.
\newblock \enquote{Large Language Models Encode Clinical Knowledge.}
\newblock \textit{Nature} Vol. 620 No. 7972 (2023): pp. 172--180.
\newblock \doi{10.1038/s41586-023-06291-2}.

\bibitem{xieApplicationTRIZInnovation2023}
Xie, Qizhi and Liu, Qiang.
\newblock \enquote{Application of {{TRIZ Innovation Method}} to {{In-Pipe Robot
  Design}}.}
\newblock \textit{Machines} Vol.~11 No.~9 (2023): p. 912.
\newblock \doi{10.3390/machines11090912}.

\bibitem{hamalainenEvaluatingLargeLanguage2023}
H{\"a}m{\"a}l{\"a}inen, Perttu, Tavast, Mikke and Kunnari, Anton.
\newblock \enquote{Evaluating {{Large Language Models}} in {{Generating
  Synthetic HCI Research Data}}: A {{Case Study}}.}
\newblock \textit{Proceedings of the 2023 {{CHI Conference}} on {{Human
  Factors}} in {{Computing Systems}}}: pp. 1--19. 2023. {ACM}, {Hamburg
  Germany}.
\newblock \doi{10.1145/3544548.3580688}.

\bibitem{changTRIZinspiredKnowledgedrivenApproach2023}
Chang, Danni, Li, Fan, Xue, Jiao and Zhang, Liqun.
\newblock \enquote{A {{TRIZ-inspired}} Knowledge-Driven Approach for
  User-Centric Smart Product-Service System: {{A}} Case Study on Intelligent
  Test Tube Rack Design.}
\newblock \textit{Advanced Engineering Informatics} Vol.~56 (2023): p. 101901.
\newblock \doi{10.1016/j.aei.2023.101901}.

\bibitem{gilani2023improving}
Gilani, Fazal and Lee, Ching-Hung.
\newblock \enquote{Improving Usability Services and User Experience of Online
  Museum Virtual Exhibition with TRIZ Theory.}
\newblock \textit{International TRIZ Future Conference}: pp. 405--416. 2023.
  Springer.

\bibitem{kim2022conceptual}
Kim, Taehoon, Lim, Hyunsu and Cho, Kyuman.
\newblock \enquote{Conceptual robot design for the automated layout of building
  structures by integrating QFD and TRIZ.}
\newblock \textit{The International Journal of Advanced Manufacturing
  Technology} Vol. 120 No.~3 (2022): pp. 1793--1804.

\bibitem{gilaniImprovingUsabilityServices2023}
Gilani, Fazal and Lee, Ching-Hung.
\newblock \enquote{Improving {{Usability Services}} and {{User Experience}} of
  {{Online Museum Virtual Exhibition}} with {{TRIZ Theory}}.}
\newblock Cavallucci, Denis, Livotov, Pavel and Brad, Stelian (eds.).
  \textit{Towards {{AI-Aided Invention}} and {{Innovation}}}.
\newblock Vol. 682. {Springer Nature Switzerland}, {Cham} (2023): pp. 405--416.
\newblock \doi{10.1007/978-3-031-42532-5_31}.

\end{thebibliography}



\appendix

\section{Prompt details}
\label{appendix: prompt details}
In this section, we showcase the prompt details used for TRIZ-GPT workflow in Table \ref{Tab: prompt strategies}. The content in brackets, such as [CASE\textunderscore DESCRIPTION], is provided by human designers.

\begin{table*}
\footnotesize
\centering
\caption{\textbf{Prompt details}}
\label{Tab: prompt strategies}
\vspace{-0.04in}
\label{tab:prompt}
\begin{tabular}{p{0.12\textwidth}|p{0.85\textwidth}} 
    \hline
    \rowcolor{lightgray} \textbf{Type} & \textbf{Prompt details} \\
    \hline
    \hline
    \multicolumn{2}{l}{\textbf{Step 1: Problem analysis}} \\
    \hline
    \multicolumn{2}{p{1\textwidth}}{\textit{You are a designer proficient in the use of TRIZ, able to skillfully apply TRIZ theory to analyze problems. When given a specific problem, you can easily extract key parameters from the problem description that can help solve the problem. Please analyze the following problem description and summarize the key problem parameters within the context in the following format: Parameter number, Parameter name, Parameter explanation.
    \newline
        [CASE\textunderscore DESCRIPTION]}} \\
    \hline
    \multicolumn{2}{l}{\textbf{Step 2: Parameters mapping}} \\
    \hline
    \multicolumn{2}{p{1\textwidth}}{\textit{You are a designer proficient in the use of TRIZ, able to skillfully apply TRIZ theory to analyze problems. When you are given a specific problem and its problem parameters, you can easily map these problem parameters to the standard TRIZ engineering parameters. In other words, for each provided problem parameter, select the appropriate corresponding parameters from the standard TRIZ engineering parameter list. 
    \newline
    Remember to tell me the number of the corresponding TRIZ engineering parameters.
    \newline
    1. Weight of Moving Object - The mass of an object that is in motion.
    \newline
    2. Weight of Stationary Object - The mass of an object that is not moving.
    \newline
    3. Length of Moving Object - The measurement from end to end of an object that is in motion.
    \newline
    ... (The complete table of TRIZ engineering parameters is omitted due to space constraints.)
    \newline
    [PROBLEM\textunderscore PARAMETERS\textunderscore SELECTED]}} \\
    \hline
    \multicolumn{2}{l}{\textbf{Step 3: Contradiction analysis}} \\
    \hline
    Basic prompt & \textit{You are a designer proficient in the use of TRIZ, able to skillfully apply TRIZ theory to analyze problems. When given TRIZ engineering parameters, identify which parameters get improved and which get worsened, then explain their contradictory relationship. List in the following structure: number, improved parameter, worsened parameter, and explanation.
    \newline
    [CASE\textunderscore DESCRIPTION]
    \newline
    [TRIZ\textunderscore PARAMETERS]
    \newline
    contradictory relations: 
    } \\
    \hline
    CoT & \textit{Let us think step by step. First, analyze if and how the input parameters get improved. Then, infer which parameters may worsen as a result of the chosen improving parameter within the question's context. Finally, please match the improved parameters with the worsened ones and delve deeper into their contradictory relationship to enhance the understanding of the problem.} \\
    \hline
    Few-shot & Due to space constraints, only one example of the learning shot is demonstrated here. Our proposed workflow contains three learning shots of similar form.
    \newline
    \textit{
    Problem Scenario: Existing muscle-powered submarines require a significant amount of effort for long-duration, high-speed travel.
    \newline
    Current State: The propulsion systems used in these submarines, such as propellers, are inefficient, leading to substantial energy loss. Although enclosed submarines are more comfortable, they consume more energy.
    \newline
    Pain Point: The power required by the submarines far exceeds the energy humans can sustainably provide over a few hours, limiting their performance.
    \newline
    Requirement: There is a need to reduce the water resistance of submarines and enhance their power efficiency to alleviate the operator's burden and increase the energy efficiency of the submarine.
    \newline
    Contradictory Relations:
    \newline
    Improving: 39. Productivity
    \newline
    Worsening: 33. Ease of Operation
    \newline
    Explanation: Enhancing the productivity of the submarine, such as increasing its speed and efficiency, may lead to a more complex operation, thereby making it harder for the operator to manage and control the submarine effectively.} \\
    \hline
    CoT \& Few-shot & \textit{Step-by-step reasoning path and three structured learning instances.} \\
    \hline
    \multicolumn{2}{l}{\textbf{Step 4: Solution reasoning}} \\
    \hline
    Basic prompt & \textit{You are a designer proficient in using TRIZ, skilled at deriving inspiration from the inventive principles of TRIZ to propose significant and feasible solutions. When presented with a problem description and inventive principles from TRIZ for guiding the solution, you are able to reason about the specific solution of the inventive principle for this problem situation. 
    \newline
    [CASE\textunderscore DESCRIPTION]
    \newline
    [INVENTIVE\textunderscore PRINCIPLES]} \\
    \hline
    CoT & \textit{Let's think step by step. First, explain the specific content of the given TRIZ inventive principles. Then, associate with how these principles were applied in the past problems to deepen the understanding of their usage methods. Finally, let's think outside the box. For the problem description, using the given principles as a guide, propose specific solutions. It is required that the solutions are consistent with scientific and engineering realities and are applicable to the given problem.} \\
    \hline
    Few-shot & Due to space constraints, only one example of the learning shot is demonstrated here. Our proposed workflow contains three learning shots of similar form.
    \newline
    \textit{
    Problem Scenario: Desktop fans are very essential in our life which can drive away the heat. 
    \newline
    Current State: Traditional desktop fans with large, rubber-coated blades have been in use for many years. 
    \newline
    Pain Point: Users are bothered by the noise generated by these fans. There's a risk of injury due to the exposed, large fan blades. Users experience discomfort due to irregular airflow patterns and occasional cold gusts. 
    \newline
    Requirement: Develop a fan that operates more quietly to minimize disturbances, ensures user safety and produces a consistent and pleasant airflow, eliminating sudden gusts of cold wind.
    \newline
    Inventive Principle: 28-Mechanics substitution
    \newline
    Concrete Solutions: By utilizing hydrodynamic propulsion technology, the submarine draws in water through a muscle-powered pump and ejects it at two points: at the front, the water adheres to the sides of the submarine due to the Coandă effect, reducing underwater resistance; at the rear, the water is ejected through channels on both sides of the hull, thereby enhancing propulsion efficiency.} \\
    \hline
    CoT \& Few-shot & \textit{Step-by-step reasoning path and three structured learning instances.} \\
    \bottomrule
    \end{tabular}
\end{table*}

\section{Models generation comparison example}
\label{appendix: comparison of GPT-4 & GPT-3.5}
We presented a representative example to illustrate the differences between the outputs of GPT-4 and GPT-3.5 as discussed in Section \ref{subsec: models eva}. Results that align with the contradiction parameters from reference \cite{gilaniImprovingUsabilityServices2023} are highlighted in light yellow.

\textbf{GPT-4 generated results with Chain-of-Thought:} 

1. Improved Parameter: Ease of Operation (33)

   Worsened Parameter: Device Complexity (36)
   
   Explanation: By improving the ease of operation, the user interface becomes more intuitive and user-friendly...

2. Improved Parameter: Adaptability or Versatility (35)

   Worsened Parameter: \hl{Stability of the Object (13)}
   
   Explanation: By improving the adaptability or versatility, the virtual exhibition can cater to a wider range of user preferences and needs...

3. Improved Parameter: Illumination Intensity (18)

   Worsened Parameter: \hl{Stability of the Object (13)}
   
   Explanation: By improving the illumination intensity, the virtual exhibition can create a more immersive and engaging environment...
   
\textbf{GPT-3.5 generated results with Chain-of-Thought:} 

1. Improved Parameter: Ease of Operation
Worsened Parameter: Object-Affected Harmful

Explanation: By improving the ease of operation, the virtual exhibition interface and experience will become more user-friendly and intuitive...

In summary, the selected example demonstrates that GPT-4 reasoned three contradiction pairs, one of which was used in the original literature. In contrast, GPT-3.5 only identified one pair of contradiction parameters and failed to reflect the original practice. Additionally, GPT-3.5 did not comply with the prompt’s requirement to specify the parameter numbers.

\section{Solution examples in case study}
\label{appendix: generated solutions and coding examples}
Some generated solutions by GPT-4 and GPT-3.5 as well as the coded results are presented in Table~\ref{tab:case_study_result}.

\begin{table*}
\footnotesize
\centering
\caption{\textbf{Examples of the generated solutions and their coding results in case study}}
\vspace{-0.04in}
 
	\label{tab:case_study_result}
	\begin{tabular}{p{0.2\textwidth}|p{0.5\textwidth}|p{0.2\textwidth}} 
		\hline
		\rowcolor{lightgray} \bf{Type}  & \bf{Concrete solutions} & \bf{Coded solutions}\\
		\hline
            \hline
            \multicolumn{2}{l}{\textbf{1-Segmentation}} \\
            \midrule
		Ground truth & \textit{In order to adapt to a small change in pipe diameter, according to inventive principle 1 (segmentation), we can divide the driven parts into rigid parts and flexible parts to adapt to the small change in pipe diameter.} & divide into rigid and flexible parts; replace rigid leg with rigid parts and flexible rubber mat \\
		  \midrule
    	GPT-4 generated solutions & \textit{The peristaltic robot's design can be segregated into independent units comprised of curved flexible segments (1-Segmentation). Each unit will contain a single motor and a flexible coupler, enabling each to grip and conform to different pipe diameters without requiring multiple motors per unit. This segmented approach not only simplifies manufacturing by standardizing the units, but it also enhances adaptability by permitting each unit to conform independently to the pipe's shape and size.} &  independent units comprised of curved flexible segments \\
		\midrule
		GPT-3.5 generated solutions & \textit{By implementing segmentation in the design of the bidirectional peristaltic robot, the traditional approach of using three motors can be restructured. Instead of one continuous motor system, the robot can be divided into segments, each with its own motion control mechanism. By segmenting the robot based on the varying pipe diameters it encounters, the need for multiple motors can be eliminated. This approach allows for a more adaptable design that can efficiently navigate through pipes of different sizes while simplifying manufacturing and control systems.} & divided into segments, each with its own motion control mechanism \\
            \midrule
            \multicolumn{2}{l}{\textbf{11-Beforehand cushioning}} \\
            \midrule
		Ground truth & \textit{According to inventive principle 11 (Beforehand cushioning), a pre-compression spring can be applied to the leg to ensure stable support with the pipe wall.} & pre-compression spring for stable support \\
		  \midrule
    	GPT-4 generated solutions & \textit{You can modify the design of the robot so that the motors are outfitted with spring-loaded mechanisms. These mechanisms can adapt to the width of different pipes by extending or contracting as needed, maintaining the robot's operations efficiency. The springs serve as a form of 'beforehand cushioning', allowing the robot to adjust automatically to different pipe diameters before beginning its work, which helps simplify the manufacturing process as fewer motors are needed. Additionally, the spring-loaded mechanisms, acting as buffers, can reduce wear and tear on the motors, thereby extending their lifespan.} & spring-loaded mechanisms adapt to pipe widths; extend or contract as needed \\
		\midrule
		GPT-3.5 generated solutions & \textit{By applying the inventive principle of "Beforehand cushioning," a unique system where the robot's movement is pre-programmed to adapt to various pipe diameters is implemented. This system anticipates the different diameters, allowing the robot's movement to adjust and cushion itself accordingly without the need for multiple motors. Through intelligent sensors and algorithms, the robot can predict the changes in pipe size beforehand, ensuring smooth and efficient movement without the complexity of additional motors.} & pre-programmed system with intelligent sensors and algorithms \\
        \bottomrule
	\end{tabular}
\end{table*}

\section{Quick test in product design}
\label{Appendix: sec_different domains}
In this section, we present the application results in product design, with the problem in~\cite{changTRIZinspiredKnowledgedrivenApproach2023} as an example.

\textbf{Original solution:}

Based on design principles 35, the shape of the tube rack can be redesigned. A rough design with a round shape (which provides comfortable and safer touch with no sharp edge and enables the rotation of test tubes for flexible movements), and the demountable construction of the rack and cover (for easier manufacturing, assembly and clean) was proposed.

\textbf{GPT-4 generated solution with few-shot strategy:}

1. Adjust the Parameter of Shape: Design the STTR with adaptive test tube slots that can mechanically adjust their shape to fit various diameters and lengths of test tubes. This feature uses retractable or expandable components within each slot, ensuring a secure hold for different types of test tubes used in medical testing. This design not only accommodates diverse tube sizes but also prevents the risk of test tube damage or misplacement during handling. 
\newline
2. Adjust the Parameter of Weight: Implement a lightweight, durable composite material for the structure of the STTR, designed to be easily portable yet stable enough to prevent accidents or spills. This material could also incorporate vibration-damping features to protect the samples from disturbances caused by nearby equipment or accidental bumps, enhancing the safety and reliability of the blood collection process in active healthcare environments.
\newline
3. Adjust the Parameter of Complexity: The Smart Test Tube Rack is designed with a color-coding mechanism. The rack and tubes could be color-coded based on various parameters (like the type of tests to be done) enhancing the visual identification and drastically reducing the chances of selecting the wrong test tubes, thereby simplifying the process.

\end{sloppypar}
\end{document}